# Investigation of competing magnetic orders and the associated spin-phonon coupling effect in quasi-2D $Cr_{1+\delta}Te_2$ ($\delta \approx 0.22$) single crystal

Gayathri V [1,*], Sathishkumar M [1], Sandip Kumar Kuila [2], Vikash Kumar [1], Abhidev B [1], Reshna Elsa Philip [1], Partha Pratim Jana [2], and Soham Manni [1,†]

[1] Department of Physics, Indian Institute of Technology Palakkad, Kerala - 678623, India
[2] Department of Chemistry, Indian Institute of Technology Kharagpur, West Bengal - 721302, India

**ABSTRACT**. Single crystals of quasi-2D chromium telluride system represented by $Cr_{1+\delta}Te_2$ ($\delta \approx 0.22$), which crystallizes in the trigonal structure with the *c*-axis as the growth direction, are synthesized by the flux method. The magnetization measurements revealed the coexistence and competition between ferromagnetic and antiferromagnetic exchange interactions due to the presence of the intercalated Cr-layers. A series of diverse magnetic transitions is exhibited by the crystal. While cooling the crystal, it undergoes a paramagnetic to antiferromagnetic transition at 190 K, followed by a transition into a ferromagnetic state at around 160 K, and a spin-canting at a lower temperature of 75 K. A possible lack of inversion symmetry in the crystal structure, along with the observance of an unusual jump and loop opening in the isothermal magnetization suggests that the crystals $Cr_{1+\delta}Te_2$ ($\delta \approx 0.22$) may host skyrmions. Furthermore, concomitant to their magnetic transitions, anomalies were observed in the derived Breit-Wigner-Fano fit parameters obtained from the asymmetric temperature-dependent Raman spectra, evidencing a strong spin-phonon coupling effect, intrinsic to the grown crystals. A strong perpendicular magnetic anisotropy along with the robust spin-phonon coupling, makes the system a promising candidate for prospective spintronics applications.

## I. INTRODUCTION

Due to the inherent layered structure, van der Waals (vdW) magnetic materials are considered to be potential candidates for spintronics applications in both bulk and two-dimensional (2D) limit [1–4]. A strong perpendicular magnetic anisotropy (PMA) is a key ingredient for observing a long-range magnetic order in the 2D limit by overcoming the restriction imposed by Mermin-Wagner theorem [5]. Recently, the chromium telluride (Cr-Te) system has gained attention due to its high PMA and above-room-temperature ferromagnetism [6,7]. The Cr-Te system represented by $Cr_{1+\delta}Te_2$ ($0 \leq \delta \leq 1$) belongs to the family of transition metal chalcogenides with $CrTe_2$ as the parent compound and Cr self-intercalation in between the vdW gap formed by the two neighbouring $CrTe_2$ layers [8,9]. As the concentration of Cr ($\delta$) in $Cr_{1+\delta}Te_2$ increases, Cr atoms get self-intercalated between the neighbouring $CrTe_2$ layers, thereby giving a quasi-2D nature to the system. The intercalated Cr-layer can modify the exchange interaction, thereby providing a platform to investigate diverse unconventional magnetic ground states with tunable properties like magnetic anisotropy, magnetic ordering temperature, non-collinear spin-textures, etc. [10].

Though $Cr_{1+\delta}Te_2$ was previously reported to crystallize in centrosymmetric structures [11–13], few recent reports observe a non-centrosymmetric structure for the crystal [14,15]. The lack of inversion symmetry results in an additional contribution to the magnetic exchange interaction due to the Dzyaloshinskii-Moriya interaction (DMI), which can subsequently stabilize topologically protected chiral spin textures called skyrmions in the crystal [16,17].

The observation of skyrmions in $Cr_{1+\delta}Te_2$ makes them a promising candidate for spintronics. Understanding the influence of the spins on the other degrees of freedom, like phonons is crucial for spintronic device fabrication [18]. The spin-phonon coupling effect is of particular interest, wherein the spin degree of freedom can be used to manipulate that of the lattice and vice-versa, which finds application in high-speed spintronic devices. Nevertheless, the spin-phonon coupling effect in $Cr_{1+\delta}Te_2$ single crystals remains poorly understood. Although there have been attempts to understand the spin-phonon coupling in $Cr_{1+\delta}Te_2$ thin films [19,20], the effect of strain due to the substrate could not be ignored and isolated, thereby necessitating the investigations on the free-standing single crystals.

In this work, single crystals of quasi 2D $Cr_{1+\delta}Te_2$ with $\delta \approx 0.22$ crystallizing in the NiAs prototype structure were grown by the flux method. The crystals were characterized for their structural properties and phase purity. The grown crystal was further studied for their magnetic ground state, magnetic field induced states, and spin-phonon coupling effect.

## II. EXPERIMENTAL METHODS

Single crystals of $Cr_{1+\delta}Te_2$ ($\delta \approx 0.22$) were grown by the flux method using Sn and excess Te as the flux. Detailed crystal growth procedure can be found in section 1 of the supplemental material [21]. Structural characterization at room temperature was recorded on a piece of the crystal plate using a Rigaku X-ray diffractometer in Bragg-Brentano geometry ($\lambda_{Cu\text{-}K_\alpha} = 1.5418$ Å) in the angular range 10-100° with a step size of 0.01°. The room temperature single crystal X-ray diffraction (SCXRD) data ($\lambda_{Mo}$-$K_\alpha$ =0.71073 Å) were collected using a BRUKER Photon II detector, integrated into the Bruker D8 Quest diffractometer. The data acquisition and integration processes were executed using Apex 5 software [22]. The precession images (Fig. 1(c)-(e)) in reciprocal space were constructed using the CrysAlisPro 42.49 software [23]. Compositional analyses were performed using energy-dispersive X-ray spectroscopy (EDS) measurements attached to the field emission scanning electron microscope from Carl Zeiss, Gemini SEM 300. Further, magnetization and heat capacity measurements were performed on the crystals to confirm the magnetic transitions in the grown crystals. Magnetization ($M$) measurements as a function of temperature ($T$) and magnetic field ($H$) in the two field orientations, i.e., the magnetic field applied parallel to the sample plate ($H||ab$) and perpendicular to the sample plate ($H||c$) were performed using a SQUID magnetometer (Model: MPMS- XL Ever Cool, Quantum Design, USA). Temperature and field dependent heat capacity measurements ($C_P$-$T$) were performed by relaxation technique in the $H||c$ orientation in the temperature range 2 K and 300 K, using a Physical Property Measurement System of Quantum Design, USA. To further investigate the possible existence of a spin-phonon coupling, temperature-dependent Raman spectroscopy measurements were performed in the temperature range 80 K to 300 K using Raman spectrophotometer from Horiba Labram HR Evo, Japan, in the backscattered geometry. An excitation wavelength of 532 nm with the choice of a low laser power of 2mW was employed for the measurements to avoid laser-induced heating of the sample. The $Cr_{1+\delta}Te_2$ crystals were exfoliated by mechanical exfoliation method using scotch tape to remove any $TeO_2$ layer, if formed, prior to the Raman spectroscopy measurements.

*gayathriv@iitpkd.ac.in
†smanni@iitpkd.ac.in

### III. RESULTS AND DISCUSSION
#### A. Structural and Compositional analyses

Fig. 1(a) shows the X-ray diffraction (XRD) pattern obtained by $(\theta$-$2\theta)$ scan on the largest flat surface of the crystal. The presence of reflections only from ($00l$) planes indicates that the growth direction of the crystal is along the $c$-axis and the $ab$-plane lies in the exposed crystal plane. A representative picture of the crystal indicating its growth direction is represented in the inset of Fig. 1(a). The $c$-lattice parameter determined using modified Bragg fit to the obtained peak positions is found to be 12.004(2) Å (see Fig S1) [24].

*gayathriv@iitpkd.ac.in
†smanni@iitpkd.ac.in

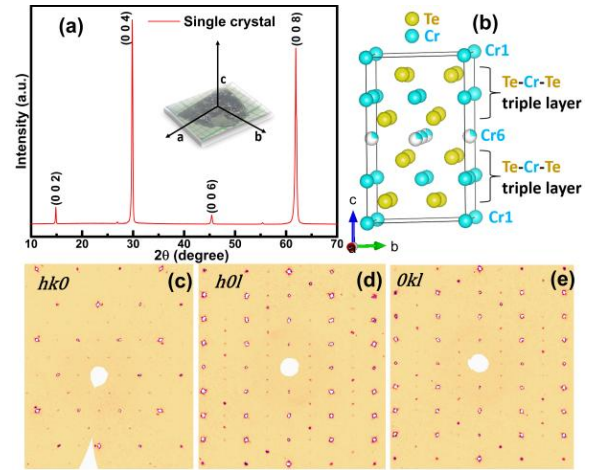

FIG 1: (a) XRD pattern ($\theta$-$2\theta$ scan) obtained by exposing the X-ray beam to a piece of flat crystal plate; (b) Unit cell: $CrTe_2$ triple layer and two different vdW gaps between two triple layers; Precession images in (c) $hk0$, (d) $0kl$, and (e) $h0l$ in the reciprocal space constructed by CrysAlisPro software.

Precession images in $hk0$, $0kl$, and $h0l$ in the reciprocal space constructed by CrysAlisPro software from the SCXRD data are depicted in Fig. 1(c)-(e), respectively. The diffraction spots obtained from the SCXRD were indexed based on the primitive trigonal lattice with the cell parameters of $a$ = 7.8233(6) Å and $c$ = 11.9818(10) Å. Interestingly, the obtained diffraction spots could be refined with both non-centrosymmetric $P3m1$ (SG No. 156) and centrosymmetric $P\bar{3}m1$ (SG No. 164) space groups using the Superflip implemented Jana 2006 software package [25,26]. The details on data collection, reduction, structure solution, and refinement are presented in Tables S1-4 in the supplemental material [21]. The acentric structure contains fourteen distinct crystallographically independent positions: six sites are occupied by chromium (Cr) atoms and the remaining eight crystallographic positions are occupied by tellurium (Te) atoms, whereas the centro-symmetric structure consists of eight distinct Wyckoff sites. The structure is viewed as the alternative arrangement of Cr and Te layers stacked perpendicular to [$00l$]. A crucial structural feature of this structure is the presence of van der Waals (vdW) gaps between the two triple layers. Notably, the vdW gaps are accommodated by fully occupied Cr1 atoms and partially occupied Cr6. However, despite these differences in vdW spacing, both the $CrTe_2$ triple-layer units remain identical. The stoichiometry of the crystal obtained from the refinement of the SCXRD data using the non-centrosymmetric $P3m1$ space group was found to be $Cr_{1.222(4)}Te_2$. The unit cell of the $Cr_{1.222(4)}Te_2$ with $CrTe_2$ triple layer and two different vdW gaps between the two triple layers is illustrated in Fig. 1(b). The stoichiometry of the crystal calculated from the atomic percentages of the constituent elements obtained from the chemical compositional analyses using EDS measurement (see section 4 of supplemental material [21]) was found to be $Cr_{1.24(6)}Te_{2.00(4)}$, which is consistent with the stoichiometry determined from the refinement of the SCXRD data. The crystallographic information files (CIF) are deposited via the joint Cambridge Crystallographic Data Centre and Fachinformationszentrum Karlsruhe deposition service (CSD: 2449820-2449821).

### B. Magnetic properties

Temperature-dependent susceptibility ($\chi$-$T$) curves for $H||c$ ($\chi_c$) and $H||ab$ ($\chi_{ab}$) in zero field cooled warming (ZFC) and field cooled cooling (FCC) modes, are shown in Fig. 2(a). By comparing the susceptibility along the two directions, $\chi_c$ is found to be an order of magnitude higher than $\chi_{ab}$, indicating a strong perpendicular magnetic anisotropy in the crystals with its easy axis of magnetization along the $c$-direction. The vertical dotted lines in Fig. 2(a) are guides to the eye corresponding to the anomaly in the plot of $d(\chi_c T)/dT$ vs $T$, indicative of the magnetic transitions, as shown in Fig. S3. The $\chi$-$T$ curves for $H$=100 Oe show the presence of three magnetic transitions: an antiferromagnetic (AFM) transition at ~190 K, followed by a ferromagnetic (FM) ordering at ~160 K, and a subsequent spin canting at ~75 K. The magnetic transitions were further confirmed by performing temperature-dependent heat capacity ($C_P$-$T$) measurements, as shown in Fig. 2(b). At high temperatures, the $C_P$ value approaches the classical Dulong-Petit limit ~ $3NR$ = 81 J/mol-K and clear anomalies are observed at the three magnetic transitions. The evolution of the transitions with an applied magnetic field ($H||c$, $H$ = 10 kOe), as obtained from $C_P$-$T$ curves, are highlighted in insets (i) and (ii) of Fig. 2(b) for transitions at 75 K, 160 K, and 190 K, respectively. A $\lambda$-like anomaly is observed at ~160 K for $H$=0 Oe (see Fig. 2(b) (ii)). This peak is further found to shift to higher temperature with the field, thereby merging with the broad AFM peak at ~180 K and resulting in a single broader peak in the temperature range 170-190 K for $H$=10 kOe. The field evolution of the $\lambda$-like peak proves the FM ordering in the crystal at $T_C$=160 K. A slope change corresponding to the spin-canted state is seen at ~75 K (Fig. 2(b) (i)). The anomalies in the $C_P$-$T$ curve for $T$>200 K are artifacts due to APIZON N grease used for mounting the crystal onto the sample holder during the measurement [27].

Soft FM hysteresis loops obtained for $T$=5, 100, and 150 K in the isothermal magnetization ($M$-$H$) measurements for $H||c$ (see Fig. 2(c) and inset (i)) further confirm the FM nature of the crystals below 160 K. The value of $M_S$ at 5 K is found to be ~2.5 $\mu_B$/Cr, consistent with the previous reports [11,13]. The competition between the AFM and FM orderings is portrayed by the obtained metamagnetic transition at $T$=170 K ($T_C$<$T$<$T_N$), as shown in the main panel of Fig. 2(c). This competition could be due to the presence of two sublattices for Cr (native and intercalated), resulting in an interlayer FM and intralayer AFM ordering [10]. However, detailed neutron scattering studies are required to confirm the nature of the observed transitions in our crystals. Additionally, a sudden drop in the magnetization value is observed at ±2 kOe as the field is swept back to zero from high fields, as highlighted in the inset (ii) of Fig. 2(c). This could be attributed to the formation of secondary magnetic phases, which were initially fully polarized and undergo a spin-reorientation at ~ ±2 kOe. Such jumps were recently reported in a non-centrosymmetric $Cr_5Te_8$ crystal in both the M-H and the topological Hall measurements, wherein the presence of the Néel skyrmions was directly observed using Lorentz transmission electron microscopy [14]. Nevertheless, the magnetic transition temperatures reported were quite different from our observations, probably due to the difference in the Cr occupancy and the associated magnetism. The absence of saturated moments in the $M$-$H$ curve up to 70 kOe for $H||ab$ (Fig. 2(d)) confirms the $c$-axis to be the easy axis of magnetization. Additionally, a loop opening in the $M$-$H$ curve around 20 kOe is observed for $H||ab$. A spin-reorientation of a coexisting secondary phase could manifest as a loop opening in the $M$-$H$ curve [28]. Hence, these sudden jumps in magnetization value seen in $H||c$ and the unusual loop opening in $H||ab$ in the isothermal magnetization can be speculated to be the signatures of skyrmion formation in the crystal. This is consistent with our SCXRD results indicating a possible lack of inversion symmetry in the crystals which could possibly host the DMI-stabilized skyrmions. Nonetheless, further detailed experimental investigations are required to ascertain this claim.

### C. Spin-phonon coupling effect

Temperature-dependent Raman spectroscopy was performed to investigate the presence of the spin-phonon coupling effect and to understand the evolution of the phonon vibrational modes as the magnetic order-to-disorder transitions happen in the crystal. $Cr_{1+\delta}Te_2$ belongs to the point-group $D_{3d}(\bar{3}m1)$ with two types of Raman active modes, non-degenerate $A_g$ mode and doubly degenerate $E_g$ mode. Here, $A_g$ and $E_g$ correspond to the out-of-plane and in-plane vibrational modes. The Raman spectra at 300 K indicate the presence of peaks at 92 cm$^{-1}$, 102 cm$^{-1}$, 123 cm$^{-1}$, and 142 cm$^{-1}$ (see Fig S4(a)). The vibrational modes at 92 cm$^{-1}$ and 102 cm$^{-1}$ were located close by and were not clearly discernible. Therefore, the thermal evolution from room temperature down to 80 K of the vibrational modes at 123 cm$^{-1}$ and 142 cm$^{-1}$ was considered for the study (see Fig. S4(b)). The Raman peaks at 123 cm$^{-1}$ and 142 cm$^{-1}$ correspond to the out-of-plane $A_{1g}$ and in-plane $E_g$ vibrational modes [29,30]. To extract the values of the peak positions and full width at half maxima (FWHM), the obtained Raman peaks were fit with different line shapes like Lorentz, Gaussian, and Voigt. However, these line shapes did not yield a good fit due to the asymmetric nature of the peaks (see Fig. S5(a) and (b)). To account for the asymmetry involved in the obtained Raman peaks, the Breit-Wigner-Fano (BWF) function given by equation (1) was used to fit the obtained peaks. Here, $I_0$ is the peak intensity, $1/q$ is the Fano parameter associated with the asymmetric line shape, $\Gamma$ is the FWHM, and $\omega_0$ is the phonon frequency in the absence of the coupling.

*gayathriv@iitpkd.ac.in
†smanni@iitpkd.ac.in

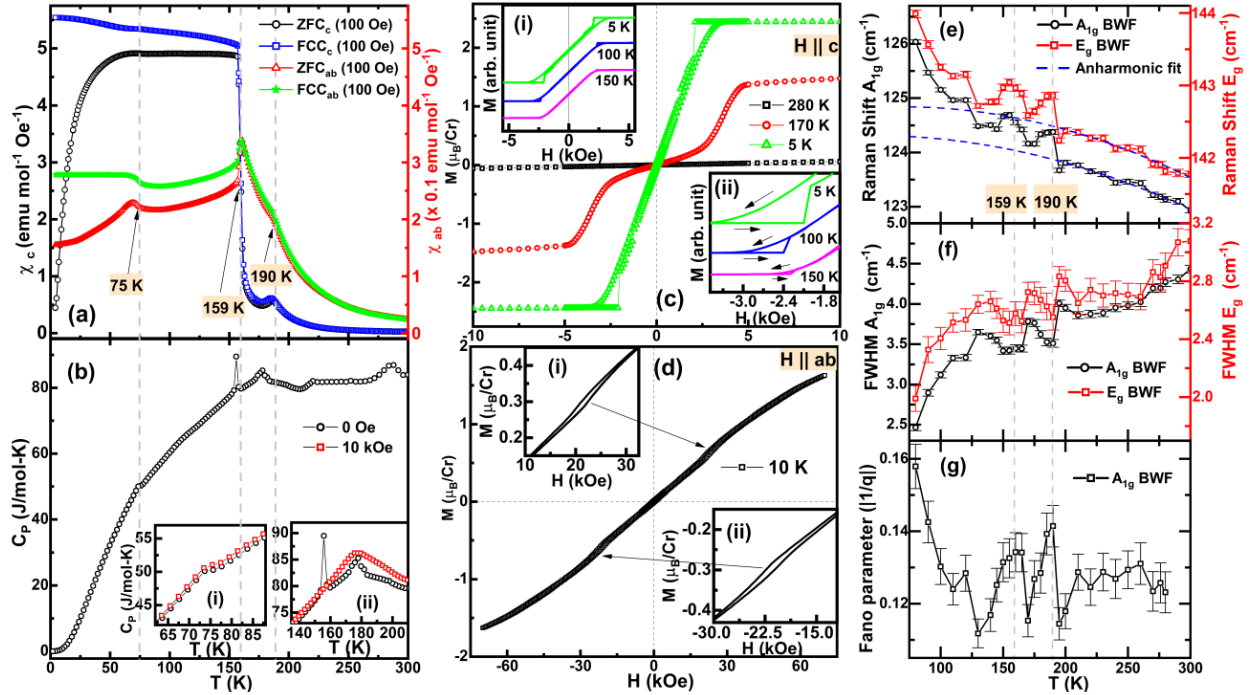

FIG 2. (a) ZFC, FCC temperature-dependent susceptibility ($\chi$-$T$) curves for $H||c$ and $H||ab$ at field $H$=100 Oe, indicating three magnetic transitions. The vertical dotted lines are guides to the eye corresponding to the anomaly in $d(\chi_c T)/dT$. (b) Temperature-dependent heat capacity ($C_P$-$T$) measurement for $H$=0 Oe confirming the magnetic transitions. Insets (i) and (ii) display the evolution of these magnetic transitions with field ($H$=10 kOe, $H||c$). Isothermal magnetization for (c) $H||c$ and (d) $H||ab$. Inset (i) and (ii) in (c) highlight the soft-ferromagnetic hysteresis loop obtained for $T<T_C$ and the jump in the $M$-$H$ curve, respectively, for $H||c$. Insets (i) and (ii) in (d) illustrate the loop opening of the $M$-$H$ curve around 20 kOe for $H||ab$. All magnetic and thermodynamic calculations were performed by considering the composition of the crystal to be $Cr_{1.22}Te_2$. Temperature evolution of (e) Raman shift, (f) FWHM, and (g) Fano parameter, as obtained from BWF fit of the Raman spectra for $A_{1g}$ and $E_g$ peaks. Blue dash lines in (e) indicate the anharmonic fit to the paramagnetic region above 200 K for both $A_{1g}$ and $E_g$ peaks.

*gayathriv@iitpkd.ac.in
†smanni@iitpkd.ac.in

$$I(\omega) = I_0 \frac{[1+2(\omega-\omega_0)/(q\Gamma)]^2}{1+4((\omega-\omega_0)^2)/\Gamma^2} \quad (1)$$

The BWF line shape was found to give a satisfactory fit to the obtained Raman peaks (Fig. S5(a) and (b)). The thermal evolution of the BWF fit parameters: Raman shift, FWHM, and Fano-parameter are plotted in Fig. 2(e)-(g), respectively. With decreasing temperature, a blue shift in the Raman peak positions is observed. The blue shift was found to be more pronounced around 190 K and 160 K. Above the magnetic ordering temperature ($T > 190$ K), i.e., in the high-temperature paramagnetic phase, the temperature-induced blue shift in the Raman peak position follows anharmonicity formalism of phonon decay with a cubic and quartic term having coefficients A and B, respectively, as given by equation (2). The cubic term pertains to the decay of a phonon of frequency $\omega_0$ into two phonons of frequency $\omega_0/2$ each, whereas the quartic term yields three phonons of frequency $\omega_0/3$ each [31]. Here, $\omega(T)$, $\omega_0$, $\hbar$, and $k_B$ are phonon frequency at temperature T, phonon frequency at 0 K, reduced Planck constant, and Boltzmann constant, respectively.

$$\omega(T) = \omega_0 + A\left[1 + \frac{2}{exp\left(\frac{\hbar\omega_0}{2k_BT}\right) - 1}\right] + B\left[1 + \frac{2}{exp\left(\frac{\hbar\omega_0}{3k_BT}\right) - 1} + \frac{3}{\left(exp\left(\frac{-\hbar\omega_0}{3k_BT}\right) - 1\right)^2}\right]$$

(2)

Anharmonicity formalism fits well for both $A_{1g}$ and $E_g$ modes down to 200 K, as shown by the blue dashed curves in Fig. 2(e). Below 200 K, the Raman shift shows deviation from this anharmonic behaviour. This deviation could arise due to the interaction of the spins with the lattice vibrations below 190 K. The blue shift in both the Raman peaks were found to be pronounced below 200 K, with an anomalous peaking around 190 K and 160 K. Interestingly, these anomalies were observed in all the BWF fit parameters (Raman shift, FWHM, and Fano-parameter) for the $A_{1g}$ mode exactly at the temperatures corresponding to the FM (160 K) and AFM (190 K) transitions. This confirms the presence of a strong spin-phonon coupling in the crystals, where the spin and lattice degrees of freedom are coupled to each other, thereby providing a tuning knob to manipulate the spin degree of freedom with that of the lattice and vice-versa. For the $A_{1g}$ mode, the value of $|1/q|$ was found to be ~0.14, which is comparable to earlier reports on other systems where Fano resonance was observed [32]. However, the value of $|1/q|$ for $E_g$ mode was found to be nearly zero, thereby reducing to a symmetric (Lorentz-type) line shape. This is confirmed from the perfect fit of the Lorentz line-shape to the deconvoluted $E_g$ peak obtained from BWF fit (see Fig. S5(c) and (d)). It is noteworthy that $E_g$ mode exhibits anomalies in the Raman shift and FWHM near the FM and AFM transitions though it is symmetric i.e., does not exhibit Fano-resonance. This could be understood by considering the perpendicular magnetic anisotropy of the grown crystals with the easy magnetization axis along the *c*-direction, resulting in the stronger coupling of the spins with the out-of-plane vibrational mode ($A_{1g}$) than with the in-plane vibrational mode ($E_g$). Hence, the presence of an intrinsic spin-phonon coupling in the free-standing crystals of $Cr_{1.22}Te_2$ was confirmed.

*gayathriv@iitpkd.ac.in
†smanni@iitpkd.ac.in

## IV. CONCLUSIONS

Single crystals of $Cr_{1.22}Te_2$ crystallizing in trigonal crystal system were successfully synthesized by the flux method. The grown crystal was found to exhibit strong perpendicular magnetic anisotropy with $\chi_c = 10\chi_{ab}$ and the growth direction of the crystal was identified to be along the *c*-axis having approximately (5 mm x 5 mm) lateral dimension. Single crystal XRD measurements revealed a possible lack of inversion symmetry in the crystals. Magnetization measurements revealed the presence of a series of three magnetic transitions: an AFM transition at 190 K, followed by a FM transition at 160 K, and a subsequent spin-canted state at 75 K. These magnetic transitions were further confirmed by heat capacity measurements. Additionally, signatures of skyrmion formation were observed in the isothermal magnetization measurements in the form of sudden jumps in the *M-H* curve for *H*||*c* and a loop opening in *H*||*ab*. These series of AFM, FM, and field induced spin canting transitions in the $Cr_{1.22}Te_2$ crystal upon lowering the temperature suggests the presence of competition and coexistence of FM, AFM, and antisymmetric exchange interactions in this system. First experimental evidence of a strong spin-phonon coupling in the free-standing crystals of $Cr_{1.22}Te_2$ was observed from the temperature-dependent Raman spectra, as illustrated by the anomalies in the BWF fit parameters near the FM and AFM transitions. Strong PMA and spin-phonon coupling, along with the possibility to host skyrmions makes this system a potential candidate for prospective spintronics applications.

## ACKNOWLEDGMENTS
Authors acknowledge CIF, IIT Palakkad and IIC, IIT Roorkee for experimental facilities. Authors thank Asha A A, Nikhil J J and Jayakumar B for discussions. SM thanks DST INSPIRE and ANRF CRG research grants for the funding provided to carry out the research. GV acknowledges Department of Science & Technology, Government of India, for financial support under WISE Post-Doctoral Fellowship programme to carry out this work.

*gayathriv@iitpkd.ac.in

†smanni@iitpkd.ac.in

*gayathriv@iitpkd.ac.in
†smanni@iitpkd.ac.in

Supplemental Material

# Investigation of competing magnetic orders and the associated spin-phonon coupling effect in quasi-2D $Cr_{1+\delta}Te_2$ ($\delta \approx 0.22$) magnetic single crystals

Gayathri V [1,*], Sathishkumar M [1], Sandip Kumar Kuila [2], Vikash Kumar [1], Abhidev B [1], Reshna Elsa Philip [1], Partha Pratim Jana [2], and Soham Manni [1,†]

[1]Department of Physics, Indian Institute of Technology Palakkad, Kerala-678623, India
[2]Department of Chemistry, Indian Institute of Technology Kharagpur, West Bengal-721302, India

1. **Single Crystal growth by self-flux method:**

Single crystals of $Cr_{1+\delta}Te_2$ ($\delta \approx 0.22$) were grown by flux method using Sn and excess Te as the flux. The precursors, Cr pieces (99.99 %, Alfa Aesar), Sn granules (99.9 %, Alfa Aesar) and Te shots (99.999 %, Alfa Aesar) in the stoichiometric ratio 1:3:10 were weighed and transferred into Canfield crucibles which were then sealed inside an 18 mm diameter evacuated quartz tube under vacuum ~$10^{-3}$ mbar. Additional Sn flux was used in order to stabilize a composition with higher Cr-content, which is otherwise hard to stabilize at room temperature, especially in the flux method of crystal growth. The temperature profile for the growth of the $Cr_{1+\delta}Te_2$ single crystals were decided by analysing the Cr-Te, Cr-Sn binary phase diagrams. The sealed ampule was initially heated from room temperature to 950 °C at the rate of 70 °C/h. It was then held there for 12 h for homogenization before cooling to 750 °C at a slow cooling rate of 3 °C/h. Finally, the ampule was centrifuged at 750 °C to separate the crystals from the molten Te flux. Shiny, prismatic plate-like single crystals with a typical dimension of 7 x 5 x 0.2 mm$^3$ were obtained in the catch crucible after the growth.

2. **Modified Bragg fit to the XRD pattern obtained by exposing the X-ray beam to a piece of flat crystal plate**

To account for the error due to crystal height in the determination of the *c*-lattice parameter from the obtained (*00l*) peaks, the modified Bragg fit given by equation S1 was used [24]. Here, h is the crystal height and R is the radius of the diffractometer. The peak position (*2θ*) was determined by fitting each of the obtained peaks with the pseudo-Voigt function. The plot of the variation of the Miller index (*l*) times $\lambda$ (1.5418 Å) as a function of the scattering angle *2θ* obtained for the single crystal for the peak positions corresponding to the (*00l*) peaks, along with the fit of modified Bragg equation, is shown in Fig. S1. The calculated *c*-lattice parameter was found to be 12.004(2) Å.

$$2c \cdot \sin\left(\theta - \frac{h}{R}\cos(\theta)\right) = l\lambda \quad \text{(S1)}$$

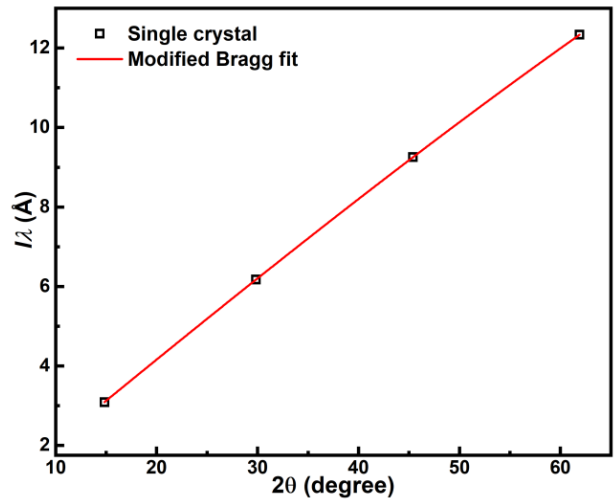

FIG S1. Determination of lattice parameter by fitting *l* times $\lambda$ as a function of the scattering angle *2θ*. The modified Bragg fit to the obtained data is shown in red curve.

*gayathriv@iitpkd.ac.in
†smanni@iitpkd.ac.in

### 3. Single crystal X-ray diffraction:

Independent refinement of Cr1 sites (1$a$) led to the site occupancy factor greater than 1, suggesting that some heavier atoms may be mixed with Cr1. Hence, the site was initially assigned as a mixed Cr/Sn (as Sn was used as a flux for sample preparation); this refinement leads to the composition $Cr_{1.204}Te_2Sn_{0.034}$. However, no Sn was traced in composition analysis, hence, the site was considered as a fully occupied Cr site (1$a$). The details on data collection, reduction, structure solution, and refinement of the SCXRD data are presented in Tables S1-4. The parameters obtained by refining as non-centrosymmetric $P3m1$ and centrosymmetric $P\bar{3}m1$ space groups are tabulated in Table S1 and S2, respectively. The coordinates, site occupancy factor, and isotropic displacement parameters corresponding to $P3m1$ and $P\bar{3}m1$ space groups are tabulated in Tables S3 and S4, respectively.

TABLE S1: Single crystal X-ray diffraction refinement in the non-centrosymmetric $P3m1$ space group

| **Chemical formula** | $Cr_{1.222}Te_2$ |
|---|---|
| $M_r$ | 318.7 |
| Crystal system, space group | Trigonal, $P3m1$ |
| Temperature (K) | 293 |
| $a, c$ (Å) | 7.8233(6), 11.9818(10) |
| $V$ (Å$^3$) | 635.09(9) |
| $Z$ | 8 |
| Radiation type | Mo $K\alpha$ |
| $\mu$ (mm$^{-1}$) | 22.01 |
| **Data collection** | |
| Diffractometer | Bruker Photon II |
| Absorption correction | multi-scan |
| No. of measured, independent and observed [$I > 3\sigma(I)$] reflections | 21316, 2415, 967 |
| $R_{int}$ | 0.076 |
| $(\sin \theta/\lambda)_{max}$ (Å$^{-1}$) | 0.837 |
| **Refinement** | |
| $R[F^2 > 3\sigma(F^2)]$, $wR(F^2)$, $S$ | 0.053, 0.165, 1.78 |
| No. of reflections | 2415 |
| No. of parameters | 62 |
| $\Delta\rho_{max}, \Delta\rho_{min}$ (e Å$^{-3}$) | 2.99, −2.15 |
| Absolute structure | 1203 of Friedel pairs used in the refinement |
| **CSD number** | 2449821 |

TABLE S2: Coordinates, site occupancy factor, and isotropic displacement parameters ($P3m1$, $hP28$, 156).

| Atom | Wyck. | Site | S.O.F. | $x/a$ | $y/b$ | $z/c$ | U [Å$^2$] |
|---|---|---|---|---|---|---|---|
| Te1 | 1$c$ | 3$m$. | 1 | 2/3 | 1/3 | 0.3688(3) | 0.0201(10) |
| Te2 | 1$b$ | 3$m$. | 1 | -2/3 | -1/3 | -0.3669(3) | 0.0280(14) |
| Te3 | 1$b$ | 3$m$. | 1 | 1/3 | 2/3 | 0.1174(3) | 0.0191(10) |
| Te4 | 1$c$ | 3$m$. | 1 | -1/3 | -2/3 | -0.1142(3) | 0.0313(15) |
| Te5 | 3$d$ | .$m$. | 1 | 0.16765(14) | 0.3353(3) | 0.3779(3) | 0.0258(9) |
| Te6 | 3$d$ | .$m$. | 1 | -0.16681(13) | -0.3336(2) | -0.3797(2) | 0.0211(8) |
| Te7 | 3$d$ | .$m$. | 1 | 0.3352(3) | 0.16762(13) | 0.1251(2) | 0.0236(8) |
| Te8 | 3$d$ | .$m$. | 1 | -0.3345(3) | -0.16725(13) | -0.1275(2) | 0.0225(8) |
| Cr1 | 1$a$ | 3$m$. | 1 | 0 | 0 | -0.0007(10) | 0.0154(10) |
| Cr2 | 1$a$ | 3$m$. | 1 | 0 | 0 | 0.2537(10) | 0.030(3) |
| Cr3 | 1$a$ | 3$m$. | 1 | 0 | 0 | -0.2558(9) | 0.019(2) |
| Cr4 | 3$d$ | .$m$. | 1 | 0.4927(4) | 0.5073(4) | 0.2478(8) | 0.0224(18) |
| Cr5 | 3$d$ | .$m$. | 1 | -0.4920(4) | -0.5080(4) | -0.2490(8) | 0.027(2) |

*gayathriv@iitpkd.ac.in
†smanni@iitpkd.ac.in

| Cr6 | 3d | .m. | 0.260(9) | 0.5045(13) | 0.4955(13) | 0.500(2) | 0.012(2) |

TABLE S3: Single crystal X-ray diffraction refinement in the centrosymmetric $P\bar{3}m1$ space group

| Chemical formula | $Cr_{1.216}Te_2$ |
|---|---|
| $M_r$ | 318.4 |
| Crystal system, space group | Trigonal, $P\bar{3}m1$ |
| Temperature (K) | 293 |
| $a$, $c$ (Å) | 7.8233(6), 11.9818(10) |
| $V$ (Å$^3$) | 635.09(9) |
| $Z$ | 8 |
| Radiation type | Mo $K\alpha$ |
| $\mu$ (mm$^{-1}$) | 21.99 |
| **Data collection** | |
| Diffractometer | Bruker Photon II |
| Absorption correction | Multi-scan |
| No. of measured, independent and observed [$I > 3\sigma(I)$] reflections | 21316, 1212, 542 |
| $R_{int}$ | 0.078 |
| $(\sin\theta/\lambda)_{max}$ (Å$^{-1}$) | 0.837 |
| **Refinement** | |
| $R[F^2 > 3\sigma(F^2)]$, $wR(F^2)$, $S$ | 0.057, 0.206, 1.76 |
| No. of reflections | 1212 |
| No. of parameters | 34 |
| $\Delta\rho_{max}$, $\Delta\rho_{min}$ (e Å$^{-3}$) | 3.58, −2.88 |
| CSD number | 2449820 |

TABLE S4: Coordinates, site occupancy factor, and isotropic displacement parameters ($P\bar{3}m1$).

| Atom | Wyck. | Site | S.O.F. | x/a | y/b | z/c | U [Å$^2$] |
|---|---|---|---|---|---|---|---|
| Te1 | 2d | 3m. | 1 | 2/3 | 1/3 | 0.36759(11) | 0.0232(4) |
| Te2 | 2d | 3m. | 1 | 1/3 | 2/3 | 0.11584(11) | 0.0241(4) |
| Te3 | 6i | .m. | 1 | 0.16733(5) | 0.33467(10) | 0.37891(7) | 0.0234(4) |
| Te4 | 6i | .m. | 1 | 0.33466(10) | 0.16733(5) | 0.12653(7) | 0.0231(4) |
| Cr1 | 1a | $\bar{3}$m. | 1 | 0 | 0 | 0 | 0.0163(8) |
| Cr2 | 2c | 3m. | 1 | 0 | 0 | 0.2554(3) | 0.0240(8) |
| Cr3 | 6i | .m. | 1 | 0.49217(12) | 0.50783(12) | 0.2482(2) | 0.0239(6) |
| Cr4 | 3f | .2/m. | 0.244(9) | 1/2 | 1/2 | 1/2 | 0.015(3) |

4. **Energy dispersive X-ray spectroscopy:**

*gayathriv@iitpkd.ac.in
†smanni@iitpkd.ac.in

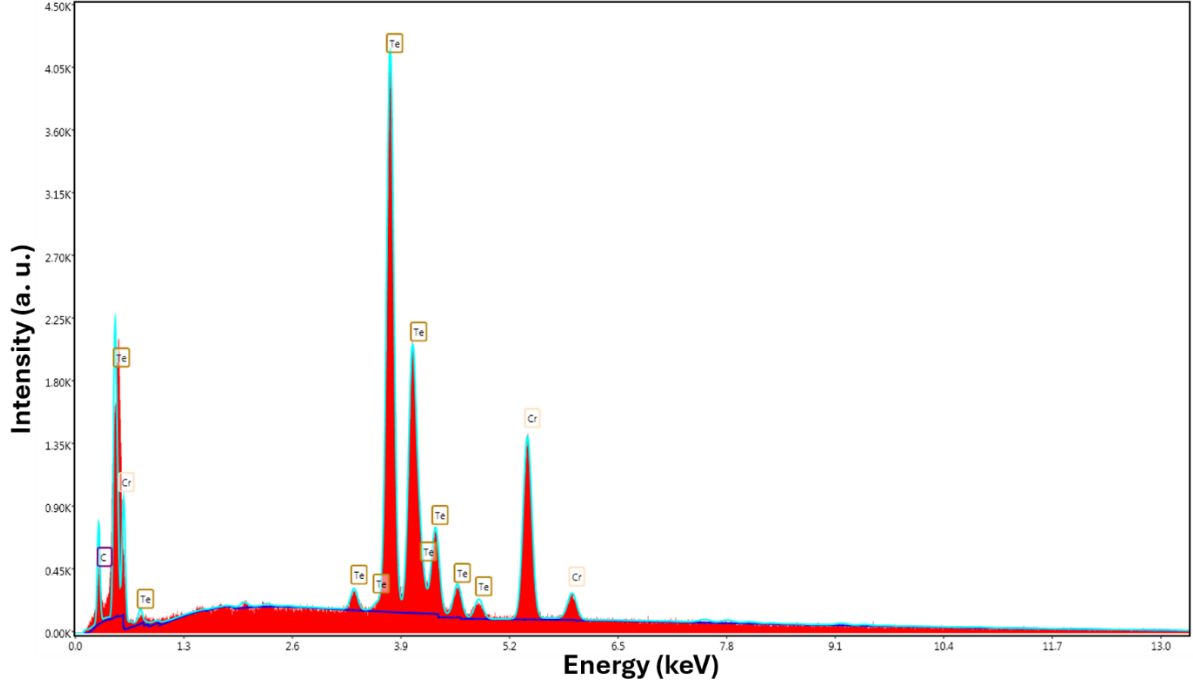

FIG S2. EDS spectra obtained for the grown crystal

Fig. S2 represents the spectra of the energy dispersive X-ray spectroscopy obtained for the crystal. The compositional analysis using EDS confirm the presence of the constituent elements (Cr and Te) in the crystal without any impurity. The atomic percentages of Cr and Te in the crystals were found to be ~ 38.22 % and 61.78 %, respectively. The stoichiometry of the sample, as calculated from the atomic percentages of the constituent elements, was found to be close to $Cr_{1.24(6)}Te_{2.00(4)}$. No elemental Sn was detected by the software. The carbon peak in the spectra is due to the contribution from the carbon tape used during the microscopy.

5. **Magnetization measurements:**

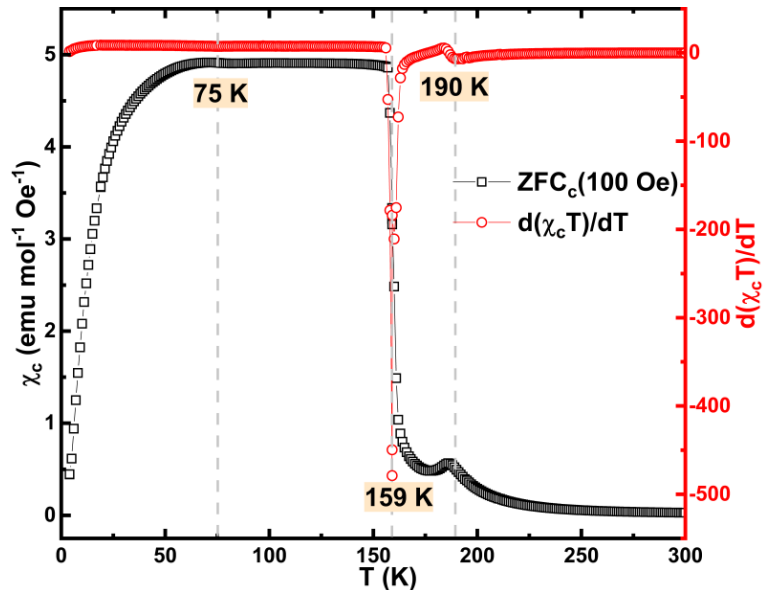

FIG S3. Determination of magnetic transition temperature from temperature dependent susceptibility ($\chi_c$ vs $T$) and $d(\chi_c T)/dT$ vs $T$ for $H||c$ direction at $H$=100 Oe in the ZFC mode.

*gayathriv@iitpkd.ac.in
†smanni@iitpkd.ac.in

Fig. S3 illustrates the determination of magnetic transition temperature from the plot of temperature dependent susceptibility for $H||c$ direction at $H=100$ Oe in the ZFC mode. The left y-axis indicates $\chi_c$ vs $T$ curve and the right y-axis indicates $d(\chi_c T)/dT$ vs $T$ curve. The anomalies at $T= 190$ K, 159 K, and 75 K in the $d(\chi_c T)/dT$ vs $T$ curve indicate the magnetic transitions corresponding to paramagnetic to antiferromagnetic, antiferromagnetic to ferromagnetic, and spin-canting.

6. **Spin-phonon coupling effect:**

Room temperature Raman spectra obtained for the crystal illustrating the observed vibrational modes is shown in Fig. S4(a). Temperature-dependent Raman spectra of the grown crystal in the temperature range of 80 K to 300 K are shown in Fig. S4(b). The prominent peaks in each of the spectra were found to be the out-plane ($A_{1g}$) and the in-plane ($E_g$) atomic vibration of the lattice as seen from the two distinct peaks at 123 cm$^{-1}$ and 142 cm$^{-1}$, respectively. The thermal evolution of these vibrational modes was studied to understand the phonon dynamics of the system below the magnetic transitions. The obtained peaks in the temperature-dependent Raman spectra were fit using different line shapes like Gaussian, Lorentz, and Voigt. However, due to the asymmetric nature of the $A_{1g}$ peak, these line shapes did not give a good fit. To account for the asymmetry associated with the $A_{1g}$ peak, the Briet-Wigner-Fano (BWF) line shape was used to fit the spectra. Representative Raman spectra for $T=300$ K and $T= 80$ K along with the fit using various line shapes, are illustrated in Fig. S5(a) and (b), respectively.

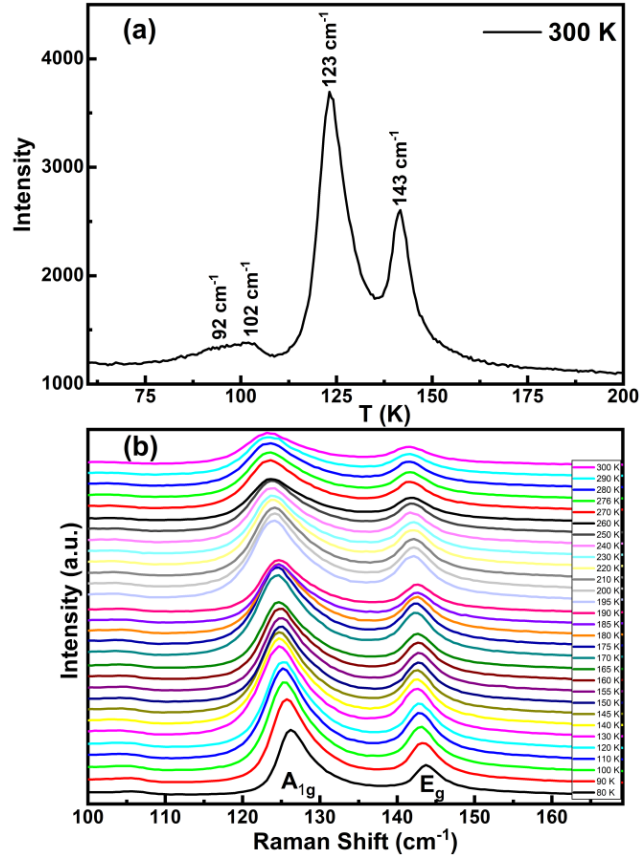

FIG S4. (a) Raman spectra of $Cr_{1.22}Te_2$ crystal at 300 K illustrating the observed vibrational modes; (b) Temperature-dependent Raman spectra of $Cr_{1.22}Te_2$ crystal illustrating the evolution of $A_{1g}$ and $E_g$ peaks as a function of temperature. Spectra obtained at different temperatures are vertically offset for clarity.

*gayathriv@iitpkd.ac.in
†smanni@iitpkd.ac.in

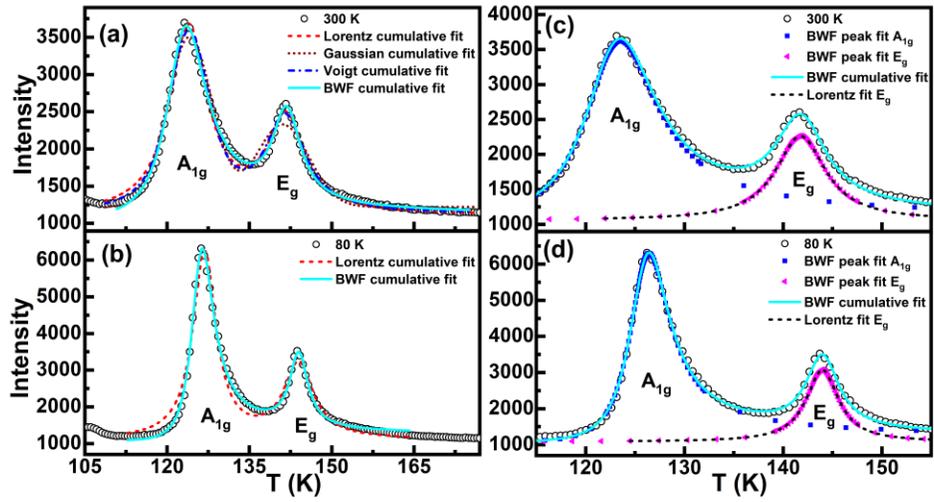

FIG S5. Zoomed-in view of the Raman spectra highlighting the asymmetric nature of the $A_{1g}$ peak at (a) 300 K and (b) 80 K. Black open circle represents the original data. The cumulative envelopes obtained from Lorentz, Gaussian, Voigt, and BWF fit to the obtained data are represented by dash, dot, dash-dot, and solid lines, respectively. The BWF fit at (c) 300 K and (d) 80 K with the deconvoluted $A_{1g}$ and $E_g$ peaks are represented by the solid square and triangles, respectively. The Lorentz fit to the deconvoluted $E_g$ peak is indicated by black dash line.

*gayathriv@iitpkd.ac.in
†smanni@iitpkd.ac.in